\documentclass[preprint,aps,prc,showpacs,preprintnumbers,superscriptaddress,amsmath,amssymb,twocolumn,10pt,floatfix]{revtex4-2}
\usepackage{dcolumn}
\usepackage{graphicx}
\usepackage{bm}
\usepackage{multirow}
\usepackage{CJK}
\usepackage[colorlinks,linkcolor=blue,anchorcolor=black,urlcolor=blue,citecolor=blue,CJKbookmarks=True]{hyperref}
\usepackage{ulem}
\usepackage{color}

\newcommand{\delete}{\bgroup\markoverwith{\textcolor{blue}{\rule[0.5ex]{2pt}{1pt}}}\ULon}

\def\dif{\mathop{}\hphantom{\mskip-\thinmuskip}\mathrm{d}}
\let\daccent\d
\let\d\relax
\newcommand\d{\ifmmode\dif\else\expandafter\daccent\fi}

\begin{document}

\begin{CJK*}{UTF8}{gbsn}
\title{PSR J$ \mathbf {0952-0607} $ and GW170817: Direct Multimessenger Constraints on Neutron Star Equation of State through a Novel Wide-ranging Correlation}
\author{L.\,Guo\,(郭兰)}
\author{Y.\,F.\,Niu\,(牛一斐)}\email{niuyf@lzu.edu.cn}
\affiliation{School of Nuclear Science and Technology, Lanzhou University, Lanzhou 730000, China}
\affiliation{Frontier Science Center for Rare isotope, Lanzhou University, Lanzhou 730000, China}
\date{\today}

\begin{abstract}
	Our knowledge about neutron star (NS) masses is renewed once again due to the recognition of the heaviest NS PSR J$ 0952-0607 $. By taking advantage of both mass observations of super massive neutron stars and the tidal deformability derived from event GW170817, a joint constraint on tidal deformability is obtained. A wide-ranging correlation between NS pressure and tidal deformability within the density range from saturation density $\rho_0$ to $5.6\rho_0$ is discovered, which directly yields a constrained NS EoS. The newly constrained EoS has a small uncertainty and a softer behavior at high densities without the inclusion of extra degrees of freedom, which shows its potential to be used as an indicator for the component of NS core. 
\end{abstract}

\maketitle
\end{CJK*}

The equation of state (EoS) of nuclear matter is essentially important for both nuclear physics\cite{sorensen2023dense} and astrophysics \cite{ozel2016masses}. However, the nuclear EoS is still poorly determined, in particular at high densities or with large isospin asymmetry. Our knowledge about the nuclear EoS mainly comes from the properties of heavy nuclei, where their density is typically limited below the nuclear saturation density $ \rho_0 = 2.8\times10^{14}~\rm{g\cdot cm}^{-3} $ with a relatively small isospin asymmetry $ \delta $. Fortunately, neutrons stars (NSs) are one of the most compact forms of matter in the universe with central densities reaching up to $ 5 $ to $ 10 $ times the nuclear saturation density $ \rho_0 $, and with very large isospin asymmetry $ \delta $ which is nearly $ 1 $, i.e., neutrons dominate the nucleonic component of NSs \cite{lattimer2004physics}, which provide natural laboratories for studying nuclear matter under extreme conditions. 

Macroscopic properties of NSs, such as masses, tidal deformabilities and radii are totally governed by the EoS of NSs \cite{PhysRev.55.364,PhysRev.55.374,hinderer2008tidal}, thus a great deal of information about the EoS of NSs can be revealed by observations on these properties. 
Among them, masses are the most widely observed thus the most informative property, especially those of super massive NSs \cite{J0952-0607,fonseca2021refined,cromartie2020relativistic,J0348+0432,J1810+1744,linares2018peering,kandel2020atmospheric}, which can efficiently constrain the NS mass limit $ M_{\rm max} $.
The recently discovered super massive NS, PSR J$ 0952-0607 $\cite{bassa2017lofar}, with a mass of $ M=2.35\pm0.17~M_{\odot} $ is the heaviest NS ever known\cite{romani2022psr}. Combining this discovery with observations of previous super massive NSs, a new probability distribution function (PDF) of NS mass limit $ M_{\rm max} $, which greatly challenges the stiffness of the neutron star EoSs, is proposed\cite{romani2022psr}.
The discovery of the binary NS merger gravitational wave (GW) event GW170817 \cite{abbott2017gw170817} also opened a new window to probe the tidal deformabilities of NSs which are one of the main observables provided by GW signals \cite{abbott2018gw170817}. It gives a constrained value of a 1.4-solar-mass NS to be $ \Lambda_{1.4} = 190_{-120}^{+390} (90\%) $. With the ongoing operation of available GW detectors and the development of the next generation detectors, tidal deformabilities are going to be determined more accurately in the future.

To utilize such observation information to constrain the NS EoS, the EoS is usually parametrized by different kinds of parametrizations, such as the Taylor-expansion parametrization \cite{li2021progress,tsang2020impact}, the spectral parametrization \cite{abbott2018gw170817,lindblom2012spectral,lindblom2010spectral,lindblom2016erratum} and the piecewise polytropic parametrization \cite{hebeler2013equation,kurkela2014constraining,annala2018gravitational,annala2022multimessenger,annala2020evidence}, or constructed by different energy density functionals \cite{lim2018neutron,malik2018gw170817,piekarewicz2019impact,tsang2019insights,nandi2019constraining}. The corresponding parameters are constrained either directly or through the Bayesian analysis by the observation information. Due to the lack of physics information or good correlations \cite{fortin2016neutron, tsang2019insights} between NS observables and EoS parameters, the uncertainties of the constrained EoS are consequently large. Furthermore, limited by the computing resources, the unavoidable cut-off makes the completeness of parameter space usually not as good as what is expected. Therefore it calls for innovative approaches to constrain the NS EoS in a more direct way. 

To achieve such a goal, the key is to explore the direct correlations between macroscopic properties of NSs and microscopic local behaviors. Such studies are rare but with one exception, where the correlation between NS pressure at twice saturation density $ p(2\rho_0) $ and the tidal deformability of a 1.4-solar-mass NS $ \Lambda_{1.4} $ was discovered \cite{lim2018neutron, tsang2019symmetry,tsang2019insights}. If this correlation were universal for a wide density range, the bridge between NS observations and the EoS would be built directly. Therefore, we will explore the universality of this correlation, and investigate the possibility to constrain the NS EoS through such a bridge.

In order to adopt more observation information into the above constraint, one also needs to build correlations between different NS observables and tidal deformability.
Our knowledge of masses of super massive NSs are relatively abundant, and recently the super massive NS, PSR J$ 0952-0607 $, updated our knowledge of NS mass limit again. The correlation between NS mass limit $ M_{\mathrm{max}} $ and the tidal deformability $ \Lambda $, although still insufficiently studied \cite{zhang2020constraints,malik2019tides}, allows us to convert the abundant mass information into tidal deformability, which can be further used to constrain the EoS. In this work, we will propose a novel approach to constrain the NS EoS directly through a wide-ranging correlation between the pressure of NS and tidal deformability, which at the same time considers the updated information of NS masses.  Furthermore, the confidence level  of the yielded EoS will be obtained through the PDF of each involved quantity, which overcomes the shortage of uncertainty analysis in the previous correlation studies. 

\begin{figure}[t]
	\includegraphics[width=0.49\textwidth]{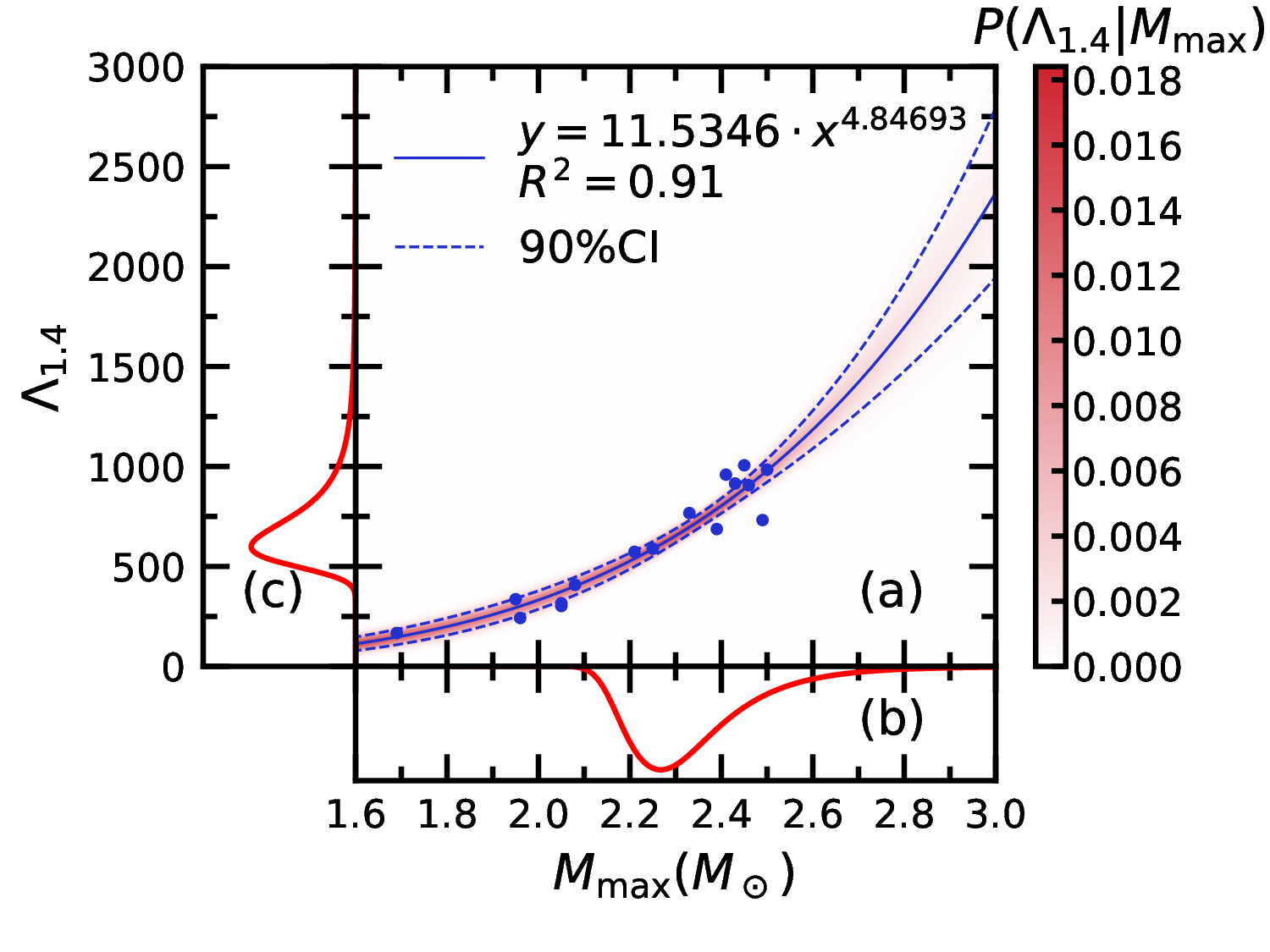}
	\caption{\label{fig-1}
		Panel (a): Tidal deformability of a 1.4-solar-mass NS $ \Lambda_{1.4} $ v.s. the NS mass limit $ M_\mathrm{max} $, given by 16 different effective interactions (blue dots). The corresponding fitted line $ \Lambda_{1.4}=11.5346  \cdot M_\mathrm{max}^{4.84693}\ $ with CoD $R^2 = 0.91 $ is plotted with solid line. The PDF of $ \Lambda_{1.4} $ for each $ M_\mathrm{max} $, $ P(\Lambda_{1.4}|M_\mathrm{max}) $, is denoted by the red shades with color bar, from which the 90\% confidence interval is obtained and its boundaries are denoted by blue dashed lines. Panel (b): The PDF of NS mass limit $P(M_\mathrm{max}|\mathrm{EM}) $ constrained by the NS mass observations taken from Ref. \cite{romani2022psr} is shown. Panel (c): The yielded PDF of $ \Lambda_{1.4} $ $ P(\Lambda_{1.4}|\mathrm{EM}) $ through the correlation between $ \Lambda_{1.4} $ and $ M_\mathrm{max} $ [see Eq. \eqref{eq-2}] is shown. 
		}
\end{figure}

Based on 16 nucleonic effective interactions (including 9 non-relativistic ones, MSL0\cite{chen2010density}, SGI\cite{van1981spin}, SIV, SV\cite{beiner1975nuclear}, SkA\cite{kohler1976skyrme}, SkM\cite{krivine1980derivation}, SLy0\cite{chabanat1995effective}, KDE0\cite{agrawal2005determination}, SAMi\cite{roca2012new}, as well as 7 relativistic ones, DD-ME2\cite{lalazissis2005new}, TW99\cite{typel1999relativistic}, PKA1\cite{long2007shell}, PKDD\cite{long2004new}, PKO1\cite{long2006density}, PKO2, PKO3\cite{long2008evolution}), we obtain 16 EoSs of NSs assuming a pure $ npe\mu $ NS core and a crust governed by the BPS+BBP model \cite{baym1971ground,baym1971neutron}. With the obtained EoS, the TOV equations \cite{PhysRev.55.364,PhysRev.55.374} and NS tidal deformabilities \cite{hinderer2008tidal} can be solved and yield the data set $ \left\{p^i(\rho);M_{\mathrm{max}}^i;\Lambda_{1.4}^i\right\} $ with $ p^i(\rho)$ the $ i $-th EoS, $ M_{\mathrm{max}}^i$ and $\Lambda_{1.4}^i$ the NS mass limit and the tidal deformability of a 1.4-solar-mass NS corresponding to the $ i $-th EoS.

\begin{figure}[t]
	\includegraphics[width=0.49\textwidth]{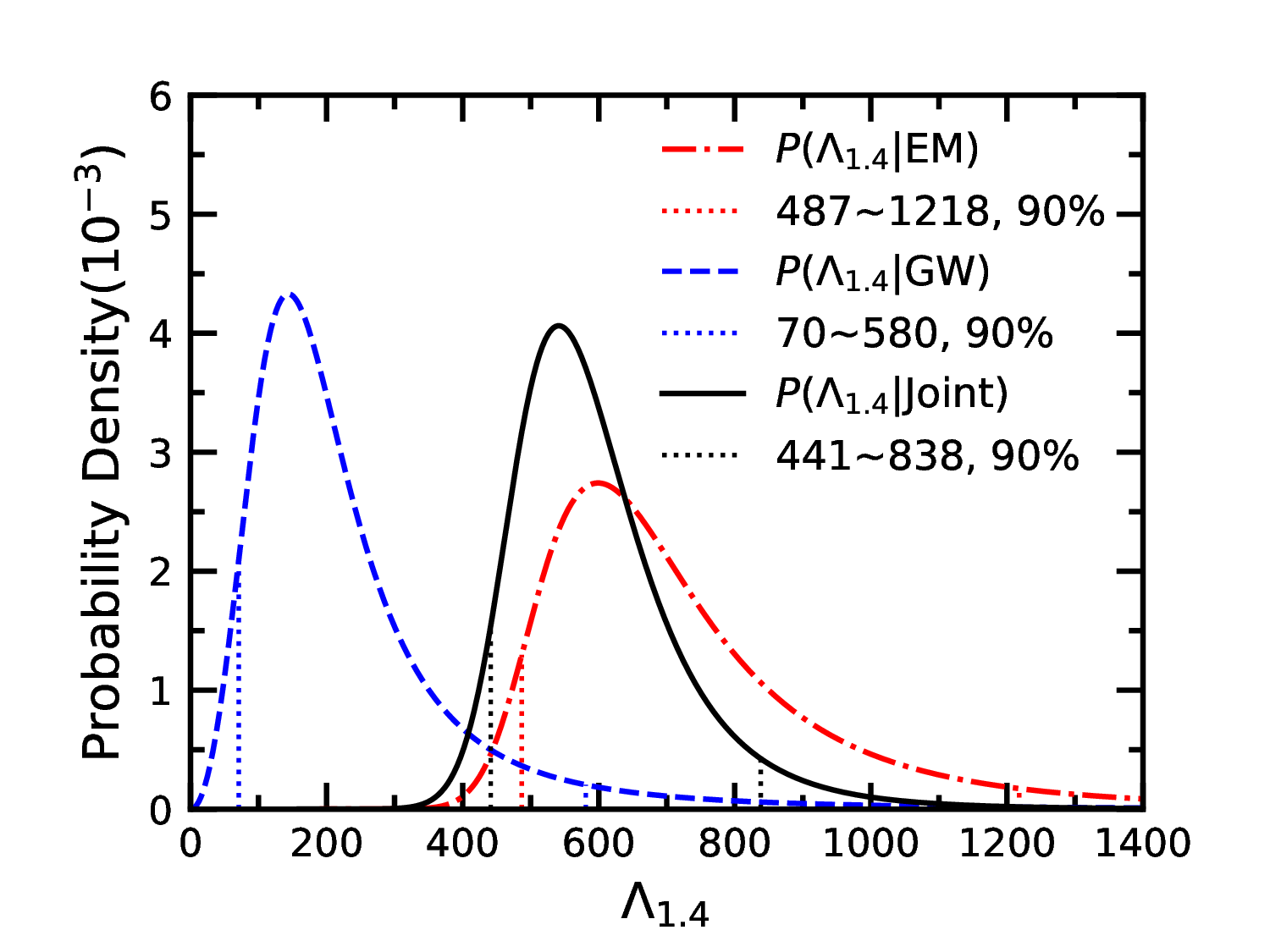}
	\caption{\label{fig-2} The PDF of the tidal deformability of a 1.4-solar-mass NS $ \Lambda_{1.4} $ constrained by different messengers. 
	$ P(\Lambda_{1.4}|\mathrm{EM}) $ constrained by super massive NSs is denoted by red dash-dotted line with the $ 90\% $ confidence interval as $\Lambda_{1.4} = 675_{-188}^{+543}$. 	$ P(\Lambda_{1.4}|\mathrm{GW}) $ constrained by GW170817 is denoted by blue dashed line which reproduces $\Lambda_{1.4}=190^{+390}_{-120}$\cite{abbott2018gw170817}. The joint distribution $ P(\Lambda_{1.4}|\mathrm{Joint}) $ combining both messengers is denoted by black solid line, with the $ 90\% $ confidence interval as $ \Lambda_{1.4} = 576^{+262}_{-135} $. 	
}
\end{figure}

As shown in the main panel of Fig. \ref{fig-1}, for the data set $ \left\{M_\mathrm{max}^i,\Lambda_{1.4}^i\right\} $, a good power-law correlation between $ M_\mathrm{max} $ and $ \Lambda_{1.4} $ is discovered since $ \left\{M_\mathrm{max}^i,\Lambda_{1.4}^i\right\} $ can be efficiently fitted using the power function $ y=a\cdot x^b $ with a coefficient of determination (CoD) up to $ R^2=0.91 $, which is 
\begin{equation}\label{eq-1}
	\Lambda_{1.4}=11.5346\cdot M_\mathrm{max}^{4.84693}.
\end{equation}
This fitting only gives the most probable $ \Lambda_{1.4} $ values for different $ M_\mathrm{max} $. However, in order to obtain the PDF of $ \Lambda_{1.4} $ at each given $ M_\mathrm{max} $, i.e.,  $P(\Lambda_{1.4}|M_\mathrm{max})$,  one can use Eq. (1) and Eq. (2) in the Supplemental material\cite{Supplemental} and the corresponding results are shown in the red shades with color bar in panel (a) of Fig. \ref{fig-1}. 

Considering the fact that the observations of masses of super massive NSs are relatively abundant compared to the poorly known tidal deformability, the discovered correlation with $ R^2>0.9 $ makes it reliable to use $ P(\Lambda_{1.4}|M_\mathrm{max}) $ as a converter from the information of $ M_\mathrm{max} $ to that of $ \Lambda_{1.4} $. 
From electromagnetic (EM) observations of super massive NSs, especially the novel heaviest NS PSR J$ 0952-0607 $, the state-of-the-art distribution about the NS mass limit $ P(M_\mathrm{max}|\mathrm{EM}) $ is extracted in Ref. \cite{romani2022psr}, which is shown in panel (b) of Fig. \ref{fig-1}. Combining this result with $ P(\Lambda_{1.4}|M_\mathrm{max}) $, the following integration
\begin{equation}\label{eq-2}
	P(\Lambda_{1.4}|\mathrm{EM})\!=\!\!\int\!\dif M_\mathrm{max}\!\cdot\! P(\Lambda_{1.4}|M_\mathrm{max})P(M_\mathrm{max}|\mathrm{EM})
\end{equation}
yields the distribution $ P(\Lambda_{1.4}|\mathrm{EM}) $, which is shown in panel (c) of Fig. \ref{fig-1} as well as in Fig. \ref{fig-2} with the red dash-dotted line. The $ 90\% $ confidence interval of this distribution gives $\Lambda_{1.4} = 675_{-188}^{+543}$. 

\begin{figure}[h]
	\includegraphics[width=0.49\textwidth]{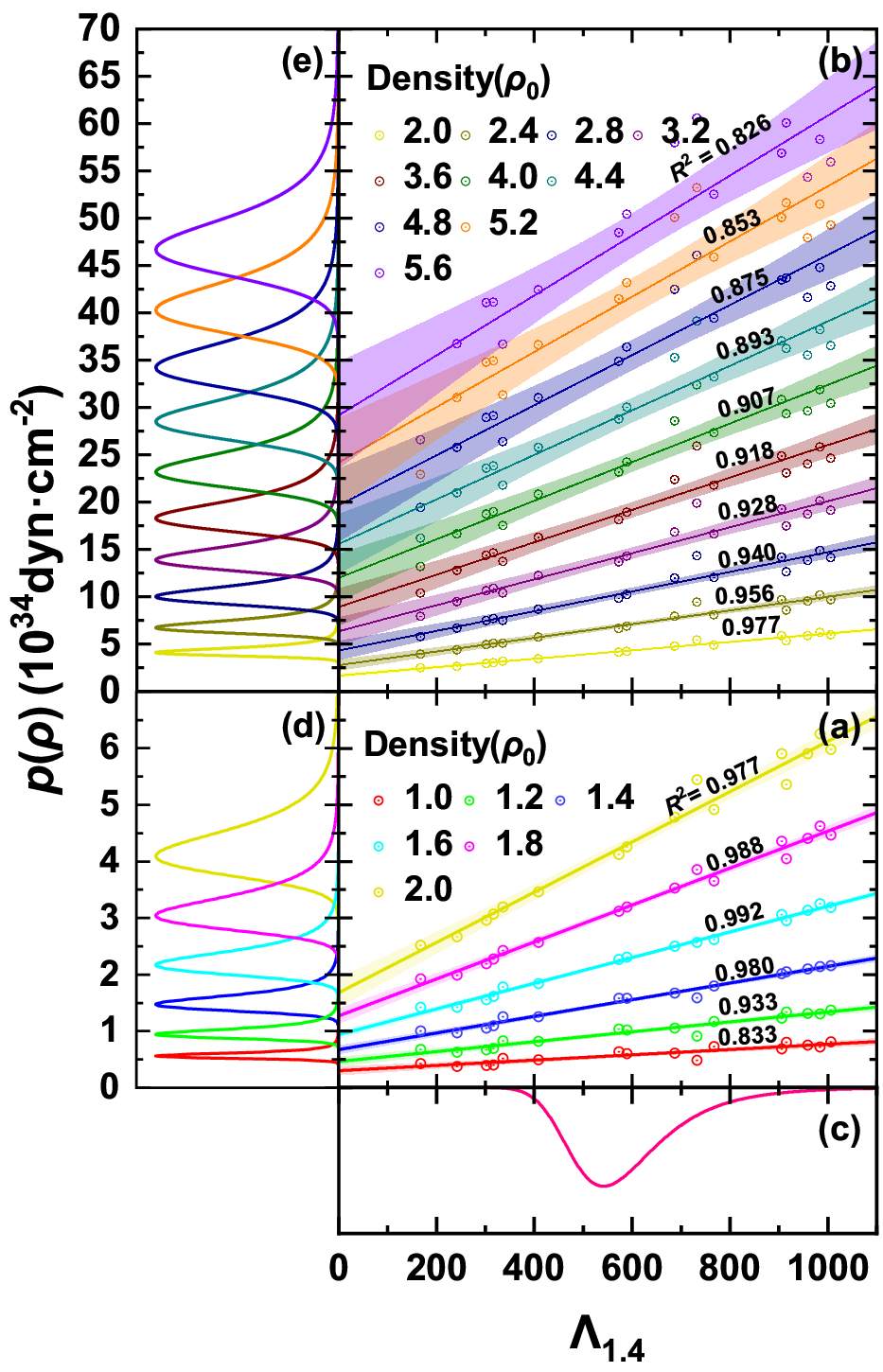}
	\caption{\label{fig-3} 
	Pressure at different densities $\rho= \rho_0 \sim 2.0 \rho_0$ [panel (a)] and $\rho= 2.0\rho_0 \sim 5.6 \rho_0 $ [panel (b)] $p(\rho) $ v.s. tidal deformability of a 1.4-solar-mass NS $ \Lambda_{1.4} $, given by 16 different effective interactions (open circles). The linear fitted lines and the corresponding $ 90\% $ confidence intervals are shown by solid lines and shaded areas, respectively, with its CoD $ R^2 $ on the top of each fitted line. The joint PDF of $ \Lambda_{1.4} $, $ P(\Lambda_{1.4}|\mathrm{Joint}) $, is shown in panel (c). The yielded PDF of pressure at each density $ P(p(\rho)|\mathrm{Joint}) $ through the correlation between $p(\rho) $ and $ \Lambda_{1.4} $ is shown in panel (d) for density range $\rho=1.0\rho_0 \sim 2.0 \rho_0$ and panel (e) for density range $\rho= 2.0\rho_0 \sim 5.6 \rho_0 $. }
\end{figure}

Besides the above yielded constraints on $ \Lambda_{1.4}$ from the observation of NS masses, the direct constraint on $ \Lambda_{1.4}$ from NS merger event GW170817 is also known, which is $ \Lambda_{1.4}=190^{+290}_{-120} $ \cite{abbott2018gw170817}. In order to get a PDF of $ \Lambda_{1.4}$ that reproduces this constraint, we follow the procedure given in Ref. \cite{kumar2019inferring} which is also shown in the Supplemental material \cite{Supplemental}, and the corresponding result is shown in Fig. \ref{fig-2} with the blue dashed line. We notice that these two PDFs of $ \Lambda_{1.4}$ are quite different. The EM-derived $ M_\mathrm{max} $ pushed higher by the newly discovered NS PSR J$ 0952-0607 $ prefers larger tidal deformabilities, resulting in a quite different PDF compared with the well-known one from GW170817. The larger tidal deformabilities preferred by NS mass limit compared to that given by GW170817  calls for more observations of NS mergers to constrain tidal deformabilities. As two independent constraints, we use the following joint distribution to combine the information from both messengers, 
\begin{equation}\label{eq-3}
	P(\Lambda_{1.4}|\mathrm{Joint})\propto P(\Lambda_{1.4}|\mathrm{EM})P(\Lambda_{1.4}|\mathrm{GW}),
\end{equation}
which gives the black solid line in Fig. \ref{fig-2}, with the $ 90\% $ confidence interval being $ \Lambda_{1.4} = 576^{+262}_{-135} $. Both bounds of this joint result are much higher than those constrained by GW170817, due to the influence of $P(\Lambda_{1.4}|\mathrm{EM})$. Due to $P(\Lambda_{1.4}|\mathrm{GW})$ from GW170817, the upper bound of $ 90\% $ confidence interval of the joint result is much lower than that of $P(\Lambda_{1.4}|\mathrm{EM})$. Therefore, the joint distribution gives a comprehensive constraint on $\Lambda_{1.4}$ with information from both messengers considered. 

\begin{figure}[t]
	\includegraphics[width=0.49\textwidth]{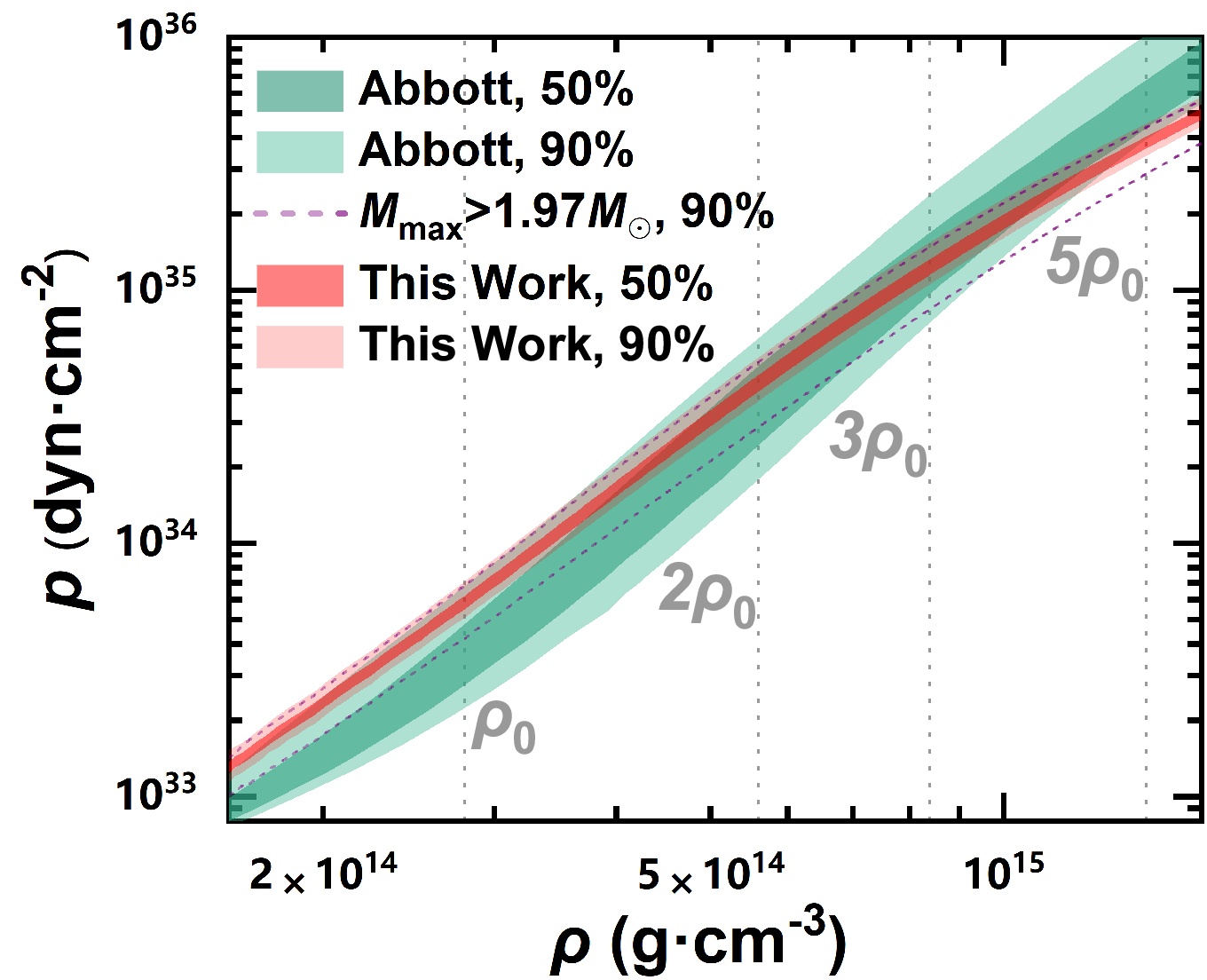}
	\caption{\label{fig-4} The NS EoS $ p(\rho) $ under different constraints. The light (dark) red shades denote the EoS constrained by $ P(\Lambda_{1.4}|\mathrm{Joint}) $ containing information from both GW event GW170817 \cite{abbott2018gw170817} and the EM observations on super massive NSs \cite{romani2022psr} with 90\% (50\%) confidence interval. By replacing the NS mass limit to a 0 distribution of $ M_\mathrm{max} \in [1.97M_\odot, 3M_\odot] $, the constrained EoS with 90\% confidence interval is given by violet dashed lines. For comparison, the constrained EoS with 90\% (50\%) confidence interval from Ref. \cite{abbott2018gw170817} is also shown by the light (dark) green shades. Some representative densities are marked by gray dotted lines.}
\end{figure}

In order to convert the observation information of $ \Lambda_{1.4} $ to the EoS, one needs to investigate correlations between the macroscopic properties of NSs and quantities in the EoS. Here, we find a surprisingly strong linear correlation between $ \Lambda_{1.4} $ and the pressure $ p(\rho) $ at each specific density in the range of $ \rho_0 $ to $ 5.6\rho_0 $ with CoDs $ R^2>0.81 $ (in the linear case $R^2=r^2$ with $r$ being the Pearson's coefficient), as shown in panel (a) and (b) of Fig. \ref{fig-3}. In particular, the correlation at density $ \rho=1.6\rho_0 $ has the largest CoD $ R^2=0.992 $. These good correlations allow us to constrain the EoS of NS directly. Following the same procedure in Fig. \ref{fig-1}, we can firstly obtain the PDF of pressure $ p(\rho) $ at a given $ \Lambda_{1.4} $, i.e., $ P(p(\rho)|\Lambda_{1.4}) $, using Eq. (1) and Eq. (3) in the Supplemental material \cite{Supplemental}. Then combining the observation information $ P(\Lambda_{1.4}|\mathrm{Joint}) $ (shown again in panel (c) of Fig. \ref{fig-3}) with the converter $ P(p(\rho)|\Lambda_{1.4}) $, the integration
\begin{equation}\label{eq-4}
	\begin{split}
		&P(p(\rho)|\mathrm{Joint})\\
		=&\int\dif \Lambda_{1.4}\cdot P(\Lambda_{1.4}|\mathrm{Joint})P(p(\rho)|\Lambda_{1.4})
	\end{split}
\end{equation}
yields the PDF of pressure $ p(\rho) $ at each specific density $ \rho $, i.e., $ P(p(\rho)|\mathrm{Joint} ) $(shown in panel (d) and (e) of Fig. \ref{fig-3}), which gives exactly the NS EoS under multimessenger constraints. 

By making $ \rho $ vary continuously, we can obtain a constrained NS EoS after connecting the bounds of 90\% (50\%) confidence intervals of $ P(p(\rho)|\mathrm{Joint}) $, shown as a light (dark) red band in Fig. \ref{fig-4}. For comparison, the constrained EoS in Ref. \cite{abbott2018gw170817}, obtained through spectral parametrization taking into account the GW170817 mass configuration and NS mass limit $ M_{\mathrm{max}}>1.97M_\odot $, is also shown with green bands.  It can be seen that our constrained EoS has a remarkably lower uncertainty compared to that of Ref. \cite{abbott2018gw170817}. At lower densities ($\rho\lesssim 1.6\rho_0$), our EoS prefers the upper bound of that in Ref. \cite{abbott2018gw170817}, but at higher densities ($\rho \gtrsim 4.0\rho_0$) our EoS prefers their lower bound. It means that our EoS shows a stiffer behavior at lower densities and a softer behavior at high densities. 

The observation information from GW170817 and NS mass limit are used to constrain the EoS in both our work and Ref. \cite{abbott2018gw170817}. However, in Ref. \cite{abbott2018gw170817}, the NS mass constraints of $ M_\mathrm{max}>1.97M_\odot $ given by the $ 1\sigma $ lower mass bound of the previous heaviest NS PSR J0348+0432 discovered in 2013 \cite{J0348+0432} is considered, while in our work the state-of-the-art $ M_\mathrm{max} $ distribution $ P(M_\mathrm{max}|\mathrm{EM}) $ given in Ref. \cite{romani2022psr} including the novel heaviest NS PSR J$ 0952-0607 $ is adopted. In order to exclude the effect of different observation information, we replace $ P(M_\mathrm{max}|\mathrm{EM}) $ with a uniform distribution $ M_\mathrm{max}\in[1.97M_\odot,3M_\odot] $ whose lower bound is the same as that of Ref. \cite{abbott2018gw170817}. Here, the upper bound is a conservative estimate of $ M_\mathrm{max} $, which has a small impact on the result due to the extremely small $ P(\Lambda_{1.4}|\mathrm{GW}) $ values at high $ \Lambda_{1.4} $ values. The corresponding $ 90\% $ confidence interval of the result after the replacement of NS mass limit is shown with violet dashed lines in Fig. \ref{fig-4}. It can be seen that the EoS with the same $ M_\mathrm{max} $ constraint as Ref. \cite{abbott2018gw170817} still shows a similar behavior as before, just with larger uncertainties. It means that although our EoS is obtained under the pure $ npe\mu $ core hypothesis, it still prefers stiffer NS EoSs at lower densities and softer ones at higher densities with the same information considered, compared with that in Ref. \cite{abbott2018gw170817}, where no hypothesis on NS components is made. It tells us the nucleonic interaction is able to give softer behavior at high densities without the inclusion of extra degrees of freedom, and this phenomenon has the potential to be used as an indicator for the component of NS core. 

Furthermore, our constrained EoS has a much smaller uncertainty than that of Ref. \cite{abbott2018gw170817} even with the same observation information considered, due to the innovative constraining method making use of the newly discovered $ p$ - $\Lambda_{1.4} $ correlation. It shows that $ p$ - $\Lambda_{1.4} $ correlation is verified not only as a quick converter but also an efficient constraint on the EoS. With the inclusion of more advanced observation information, the uncertainty of the EoS is further reduced from violet dashed lines to light red region, which shows the importance of new observations, and the new observation information could be efficiently transferred into the EoS through our constraining method. Therefore, our method could serve as a useful converter between future advances in observation of NSs and the EoS. 

In summary, we proposed a novel method to constrain the NS EoS through the linear correlation between the pressure of NS at each density and the tidal deformability of a 1.4-solar-mass NS based on various density functionals including both relativistic and nonrelativistic ones. Apart from the tidal deformability constraints from GW170817, the observation information of NS mass limit, especially the novel NS mass limit given by NS PSR J$ 0952-0607 $, is also transferred to the constraints on tidal deformability through the correlation between them. With the efficient constraining method and most advanced observation information of NSs, our yielded EoS has a very small uncertainty, which shows our method could serve as a useful converter between observations of NSs and the EoS. The newly constrained EoS shows a stiffer behavior at lower densities and a softer behavior at higher densities compared to that of Ref. \cite{abbott2018gw170817} without the inclusion of extra degrees of freedom, which shows its potential to be used as an indicator for the component of NS core. 

With this adaptable new approach, future development of observation of NSs can be directly converted to the constraints on the NS EoS, which will further deepen our understanding of the EoS as well as the component of NS.  

\section*{Acknowledgments} 
 This work is supported by the National Key Research and Development (R\&D) Program under Grant No. 2021YFA1601500 and the National Natural Science Foundation of China under Grant No. 12075104.

\bibliographystyle{unsrt}

\begin{thebibliography}{10}
	
	\bibitem{sorensen2023dense}
	Agnieszka Sorensen, Kshitij Agarwal, Kyle~W Brown, Zbigniew Chajecki, Pawe{\l}
	Danielewicz, Christian Drischler, Stefano Gandolfi, Jeremy~W Holt, Matthias
	Kaminski, Che-Ming Ko, et~al.
	\newblock Dense nuclear matter equation of state from heavy-ion collisions.
	\newblock {\em Progress in Particle and Nuclear Physics}, page 104080, 2023.
	
	\bibitem{ozel2016masses}
	Feryal {\"O}zel and Paulo Freire.
	\newblock Masses, radii, and the equation of state of neutron stars.
	\newblock {\em Annual Review of Astronomy and Astrophysics}, 54:401--440, 2016.
	
	\bibitem{lattimer2004physics}
	James~M Lattimer and Maddappa Prakash.
	\newblock The physics of neutron stars.
	\newblock {\em science}, 304(5670):536--542, 2004.
	
	\bibitem{PhysRev.55.364}
	Richard~C. Tolman.
	\newblock Static solutions of einstein's field equations for spheres of fluid.
	\newblock {\em Phys. Rev.}, 55:364--373, Feb 1939.
	
	\bibitem{PhysRev.55.374}
	J.~R. Oppenheimer and G.~M. Volkoff.
	\newblock On massive neutron cores.
	\newblock {\em Phys. Rev.}, 55:374--381, Feb 1939.
	
	\bibitem{hinderer2008tidal}
	Tanja Hinderer.
	\newblock Tidal love numbers of neutron stars.
	\newblock {\em The Astrophysical Journal}, 677(2):1216, 2008.
	
	\bibitem{J0952-0607}
	CG~Bassa, Z~Pleunis, JWT Hessels, EC~Ferrara, RP~Breton, NV~Gusinskaia,
	VI~Kondratiev, S~Sanidas, L~Nieder, CJ~Clark, et~al.
	\newblock Lofar discovery of the fastest-spinning millisecond pulsar in the
	galactic field.
	\newblock {\em The Astrophysical Journal Letters}, 846(2):L20, 2017.
	
	\bibitem{fonseca2021refined}
	Emmanuel Fonseca, H~Thankful Cromartie, Timothy~T Pennucci, Paul~S Ray, A~Yu
	Kirichenko, Scott~M Ransom, Paul~B Demorest, Ingrid~H Stairs, Zaven
	Arzoumanian, Lucas Guillemot, et~al.
	\newblock Refined mass and geometric measurements of the high-mass psr j0740+
	6620.
	\newblock {\em The Astrophysical Journal Letters}, 915(1):L12, 2021.
	
	\bibitem{cromartie2020relativistic}
	H~Thankful Cromartie, Emmanuel Fonseca, Scott~M Ransom, Paul~B Demorest, Zaven
	Arzoumanian, Harsha Blumer, Paul~R Brook, Megan~E DeCesar, Timothy Dolch,
	Justin~A Ellis, et~al.
	\newblock Relativistic shapiro delay measurements of an extremely massive
	millisecond pulsar.
	\newblock {\em Nature Astronomy}, 4(1):72--76, 2020.
	
	\bibitem{J0348+0432}
	John Antoniadis, Paulo~CC Freire, Norbert Wex, Thomas~M Tauris, Ryan~S Lynch,
	Marten~H Van~Kerkwijk, Michael Kramer, Cees Bassa, Vik~S Dhillon, Thomas
	Driebe, et~al.
	\newblock A massive pulsar in a compact relativistic binary.
	\newblock {\em Science}, 340(6131):1233232, 2013.
	
	\bibitem{J1810+1744}
	Roger~W Romani, D~Kandel, Alexei~V Filippenko, Thomas~G Brink, and WeiKang
	Zheng.
	\newblock Psr j1810+ 1744: companion darkening and a precise high neutron star
	mass.
	\newblock {\em The Astrophysical Journal Letters}, 908(2):L46, 2021.
	
	\bibitem{linares2018peering}
	Manuel Linares, Tariq Shahbaz, and Jorge Casares.
	\newblock Peering into the dark side: Magnesium lines establish a massive
	neutron star in psr j2215+ 5135.
	\newblock {\em The Astrophysical Journal}, 859(1):54, 2018.
	
	\bibitem{kandel2020atmospheric}
	D~Kandel and Roger~W Romani.
	\newblock Atmospheric circulation on black widow companions.
	\newblock {\em The Astrophysical Journal}, 892(2):101, 2020.
	
	\bibitem{bassa2017lofar}
	CG~Bassa, Z~Pleunis, JWT Hessels, EC~Ferrara, RP~Breton, NV~Gusinskaia,
	VI~Kondratiev, S~Sanidas, L~Nieder, CJ~Clark, et~al.
	\newblock Lofar discovery of the fastest-spinning millisecond pulsar in the
	galactic field.
	\newblock {\em The Astrophysical Journal Letters}, 846(2):L20, 2017.
	
	\bibitem{romani2022psr}
	Roger~W Romani, D~Kandel, Alexei~V Filippenko, Thomas~G Brink, and WeiKang
	Zheng.
	\newblock Psr j0952- 0607: The fastest and heaviest known galactic neutron
	star.
	\newblock {\em The Astrophysical Journal Letters}, 934(2):L18, 2022.
	
	\bibitem{abbott2017gw170817}
	Benjamin~P Abbott, Rich Abbott, TDea Abbott, Fausto Acernese, Kendall Ackley,
	Carl Adams, Thomas Adams, Paolo Addesso, RX~Adhikari, Vaishali~B Adya, et~al.
	\newblock Gw170817: observation of gravitational waves from a binary neutron
	star inspiral.
	\newblock {\em Physical review letters}, 119(16):161101, 2017.
	
	\bibitem{abbott2018gw170817}
	Benjamin~P Abbott, Richard Abbott, TD~Abbott, F~Acernese, K~Ackley, C~Adams,
	T~Adams, P~Addesso, Rana~X Adhikari, Vaishali~B Adya, et~al.
	\newblock Gw170817: Measurements of neutron star radii and equation of state.
	\newblock {\em Physical review letters}, 121(16):161101, 2018.
	
	\bibitem{li2021progress}
	Bao-An Li, Bao-Jun Cai, Wen-Jie Xie, and Nai-Bo Zhang.
	\newblock Progress in constraining nuclear symmetry energy using neutron star
	observables since gw170817.
	\newblock {\em Universe}, 7(6):182, 2021.
	
	\bibitem{tsang2020impact}
	CY~Tsang, MB~Tsang, Pawel Danielewicz, WG~Lynch, and FJ~Fattoyev.
	\newblock Impact of the neutron-star deformability on equation of state
	parameters.
	\newblock {\em Physical Review C}, 102(4):045808, 2020.
	
	\bibitem{lindblom2012spectral}
	Lee Lindblom and Nathaniel~M Indik.
	\newblock Spectral approach to the relativistic inverse stellar structure
	problem.
	\newblock {\em Physical Review D}, 86(8):084003, 2012.
	
	\bibitem{lindblom2010spectral}
	Lee Lindblom.
	\newblock Spectral representations of neutron-star equations of state.
	\newblock {\em Physical Review D}, 82(10):103011, 2010.
	
	\bibitem{lindblom2016erratum}
	Lee Lindblom and Nathaniel~M Indik.
	\newblock Erratum: Spectral approach to the relativistic inverse stellar
	structure problem. ii [phys. rev. d 89, 064003 (2014)].
	\newblock {\em Physical Review D}, 93(12):129903(E), 2016.
	
	\bibitem{hebeler2013equation}
	K~Hebeler, JM~Lattimer, Christopher~J Pethick, and A~Schwenk.
	\newblock Equation of state and neutron star properties constrained by nuclear
	physics and observation.
	\newblock {\em The Astrophysical Journal}, 773(1):11, 2013.
	
	\bibitem{kurkela2014constraining}
	Aleksi Kurkela, Eduardo~S Fraga, J{\"u}rgen Schaffner-Bielich, and Aleksi
	Vuorinen.
	\newblock Constraining neutron star matter with quantum chromodynamics.
	\newblock {\em The Astrophysical Journal}, 789(2):127, 2014.
	
	\bibitem{annala2018gravitational}
	Eemeli Annala, Tyler Gorda, Aleksi Kurkela, and Aleksi Vuorinen.
	\newblock Gravitational-wave constraints on the neutron-star-matter equation of
	state.
	\newblock {\em Physical review letters}, 120(17):172703, 2018.
	
	\bibitem{annala2022multimessenger}
	Eemeli Annala, Tyler Gorda, Evangelia Katerini, Aleksi Kurkela, Joonas
	N{\"a}ttil{\"a}, Vasileios Paschalidis, and Aleksi Vuorinen.
	\newblock Multimessenger constraints for ultradense matter.
	\newblock {\em Physical Review X}, 12(1):011058, 2022.
	
	\bibitem{annala2020evidence}
	Eemeli Annala, Tyler Gorda, Aleksi Kurkela, Joonas N{\"a}ttil{\"a}, and Aleksi
	Vuorinen.
	\newblock Evidence for quark-matter cores in massive neutron stars.
	\newblock {\em Nature Physics}, 16(9):907--910, 2020.
	
	\bibitem{lim2018neutron}
	Yeunhwan Lim and Jeremy~W Holt.
	\newblock Neutron star tidal deformabilities constrained by nuclear theory and
	experiment.
	\newblock {\em Physical Review Letters}, 121(6):062701, 2018.
	
	\bibitem{malik2018gw170817}
	Tuhin Malik, N~Alam, M~Fortin, C~Provid{\^e}ncia, BK~Agrawal, TK~Jha, Bharat
	Kumar, and SK~Patra.
	\newblock Gw170817: Constraining the nuclear matter equation of state from the
	neutron star tidal deformability.
	\newblock {\em Physical Review C}, 98(3):035804, 2018.
	
	\bibitem{piekarewicz2019impact}
	J~Piekarewicz and FJ~Fattoyev.
	\newblock Impact of the neutron star crust on the tidal polarizability.
	\newblock {\em Physical Review C}, 99(4):045802, 2019.
	
	\bibitem{tsang2019insights}
	CY~Tsang, MB~Tsang, Pawel Danielewicz, FJ~Fattoyev, and WG~Lynch.
	\newblock Insights on skyrme parameters from gw170817.
	\newblock {\em Physics Letters B}, 796:1--5, 2019.
	
	\bibitem{nandi2019constraining}
	Rana Nandi, Prasanta Char, and Subrata Pal.
	\newblock Constraining the relativistic mean-field model equations of state
	with gravitational wave observations.
	\newblock {\em Physical Review C}, 99(5):052802(R), 2019.
	
	\bibitem{fortin2016neutron}
	M~Fortin, C~Provid{\^e}ncia, Ad~R Raduta, F~Gulminelli, JL~Zdunik, P~Haensel,
	and M~Bejger.
	\newblock Neutron star radii and crusts: uncertainties and unified equations of
	state.
	\newblock {\em Physical Review C}, 94(3):035804, 2016.
	
	\bibitem{tsang2019symmetry}
	MB~Tsang, WG~Lynch, P~Danielewicz, and CY~Tsang.
	\newblock Symmetry energy constraints from gw170817 and laboratory experiments.
	\newblock {\em Physics Letters B}, 795:533--536, 2019.
	
	\bibitem{zhang2020constraints}
	Yingxun Zhang, Min Liu, Cheng-Jun Xia, Zhuxia Li, and Subrata~Kumar Biswal.
	\newblock Constraints on the symmetry energy and its associated parameters from
	nuclei to neutron stars.
	\newblock {\em Physical Review C}, 101(3):034303, 2020.
	
	\bibitem{malik2019tides}
	Tuhin Malik, BK~Agrawal, JN~De, SK~Samaddar, C~Provid{\^e}ncia, C~Mondal, and
	TK~Jha.
	\newblock Tides in merging neutron stars: Consistency of the gw170817 event
	with experimental data on finite nuclei.
	\newblock {\em Physical Review C}, 99(5):052801(R), 2019.
	
	\bibitem{chen2010density}
	Lie-Wen Chen, Che~Ming Ko, Bao-An Li, and Jun Xu.
	\newblock Density slope of the nuclear symmetry energy from the neutron skin
	thickness of heavy nuclei.
	\newblock {\em Physical Review C}, 82(2):024321, 2010.
	
	\bibitem{van1981spin}
	Nguyen Van~Giai and H~Sagawa.
	\newblock Spin-isospin and pairing properties of modified skyrme interactions.
	\newblock {\em Physics Letters B}, 106(5):379--382, 1981.
	
	\bibitem{beiner1975nuclear}
	M~Beiner, H~Flocard, Nguyen Van~Giai, and Phu Quentin.
	\newblock Nuclear ground-state properties and self-consistent calculations with
	the skyrme interaction:(i). spherical description.
	\newblock {\em Nuclear Physics A}, 238(1):29--69, 1975.
	
	\bibitem{kohler1976skyrme}
	HS~K{\"o}hler.
	\newblock Skyrme force and the mass formula.
	\newblock {\em Nuclear Physics A}, 258(2):301--316, 1976.
	
	\bibitem{krivine1980derivation}
	H~Krivine, J~Treiner, and O~Bohigas.
	\newblock Derivation of a fluid-dynamical lagrangian and electric giant
	resonances.
	\newblock {\em Nuclear Physics A}, 336(2):155--184, 1980.
	
	\bibitem{chabanat1995effective}
	Eric Chabanat.
	\newblock Effective interactions for extreme isospin conditions.
	\newblock Technical report, Lyon-1 Univ., 1995.
	
	\bibitem{agrawal2005determination}
	BK~Agrawal, S~Shlomo, and V~Kim Au.
	\newblock Determination of the parameters of a skyrme type effective
	interaction using the simulated annealing approach.
	\newblock {\em Physical Review C}, 72(1):014310, 2005.
	
	\bibitem{roca2012new}
	X~Roca-Maza, G~Col{\`o}, and H~Sagawa.
	\newblock New skyrme interaction with improved spin-isospin properties.
	\newblock {\em Physical Review C}, 86(3):031306(R), 2012.
	
	\bibitem{lalazissis2005new}
	GA~Lalazissis, Tamara Nik{\v{s}}i{\'c}, Dario Vretenar, and Peter Ring.
	\newblock New relativistic mean-field interaction with density-dependent
	meson-nucleon couplings.
	\newblock {\em Physical Review C}, 71(2):024312, 2005.
	
	\bibitem{typel1999relativistic}
	S~Typel and HH~Wolter.
	\newblock Relativistic mean field calculations with density-dependent
	meson-nucleon coupling.
	\newblock {\em Nuclear Physics A}, 656(3-4):331--364, 1999.
	
	\bibitem{long2007shell}
	WenHui Long, Hiroyuki Sagawa, Nguyen Van~Giai, and Jie Meng.
	\newblock Shell structure and $\rho$-tensor correlations in density dependent
	relativistic hartree-fock theory.
	\newblock {\em Physical Review C}, 76(3):034314, 2007.
	
	\bibitem{long2004new}
	Wenhui Long, Jie Meng, Nguyen Van~Giai, and Shan-Gui Zhou.
	\newblock New effective interactions in relativistic mean field theory with
	nonlinear terms and density-dependent meson-nucleon coupling.
	\newblock {\em Physical Review C}, 69(3):034319, 2004.
	
	\bibitem{long2006density}
	Wen-Hui Long, Nguyen Van~Giai, and Jie Meng.
	\newblock Density-dependent relativistic hartree--fock approach.
	\newblock {\em Physics Letters B}, 640(4):150--154, 2006.
	
	\bibitem{long2008evolution}
	WenHui Long, Hiroyuki Sagawa, Jie Meng, and Nguyen Van~Giai.
	\newblock Evolution of nuclear shell structure due to the pion exchange
	potential.
	\newblock {\em Europhysics Letters}, 82(1):12001, 2008.
	
	\bibitem{baym1971ground}
	Gordon Baym, Christopher Pethick, and Peter Sutherland.
	\newblock The ground state of matter at high densities: equation of state and
	stellar models.
	\newblock {\em The Astrophysical Journal}, 170:299, 1971.
	
	\bibitem{baym1971neutron}
	Gordon Baym, Hans~A Bethe, and Christopher~J Pethick.
	\newblock Neutron star matter.
	\newblock {\em Nuclear Physics A}, 175(2):225--271, 1971.
	
	\bibitem{Supplemental}
	See Supplemental Material for details. It presents methods to obtain the PDF of
	fitted $ \hat y $ at each given $ x $ for both linear and power-law model,
	and the detailed method to reproduce the $ \Lambda_{1.4}=190^{+290}_{-120} $
	result and give $ P(\Lambda_{1.4}|\mathrm{GW}) $, which includes extra Refs.
	\cite{donaldson1987computational,wooldridge2016introductory}.
	
	\bibitem{kumar2019inferring}
	Bharat Kumar and Philippe Landry.
	\newblock Inferring neutron star properties from gw170817 with universal
	relations.
	\newblock {\em Physical Review D}, 99(12):123026, 2019.
	
	\bibitem{donaldson1987computational}
	Janet~R Donaldson and Robert~B Schnabel.
	\newblock Computational experience with confidence regions and confidence
	intervals for nonlinear least squares.
	\newblock {\em Technometrics}, 29(1):67--82, 1987.
	
	\bibitem{wooldridge2016introductory}
	Jeffrey~M Wooldridge.
	\newblock {\em Introductory econometrics: A modern approach}.
	\newblock 2016.
	
\end{thebibliography}

\end{document}